\documentclass[prb,twocolumn,showpacs,preprintnumbers,amsmath,amssymb]{revtex4}

\usepackage{amsmath,amssymb,amsfonts} 	
\usepackage{graphics}			
\usepackage[dvips]{graphicx}
\usepackage[latin1]{inputenc}

\begin{document}

\title{Inelastic scattering in a local polaron model with quadratic coupling to bosons}

\author{Thomas Olsen}
\email{tolsen@fysik.dtu.dk}

\affiliation{Danish National Research Foundation's Center of Individual
Nanoparticle Functionality (CINF),
	Department of Physics,
	Technical University of Denmark,
	DK--2800 Kongens Lyngby,
	Denmark}

\date{\today}

\begin{abstract}
We calculate the inelastic scattering probabilities in the wide band limit of a local polaron model with quadratic coupling to bosons.
The central object is a two-particle Green function which is calculated exactly using a purely algebraic approach. Compared with the usual linear interaction term a quadratic interaction term gives higher probabilities for inelastic scattering involving a large number of bosons. As an application we consider the problem hot electron mediated energy transfer at surfaces and use the delta self-consistent field extension of density functional theory to calculate and compare coupling parameters and probabilities for exciting different vibrational modes of CO adsorbed on a Cu(100) surface.
\end{abstract}
\maketitle

\section{Introduction}
The local polaron model describes a localized electronic state which is coupled to a boson field. The local state is then assumed to be hybridized with a continuum of delocalized states and are thus not an eigenstate of the electronic part of the Hamiltonian. 

One of the first applications of the model was the coupling of plasmons to core holes\cite{langreth} and valence holes\cite{cini} in metals. The boson field then represents the plasmons which can be excited by the introduction of a structureless core-hole or a valence-hole which may be hybridized with metallic states. Plasmon excitation spectra are typically measured using Electron Energy Loss Spectroscopy (EELS) where the energy loss of highly energetic electrons are measured after transmission through a metallic film. A similar application is that of deep-level spectroscopy\cite{almbladh} where the boson field represents a phonon system which can be excited by the introduction of a core-hole. Hybridization of the core-hole is then introduced to capture the degeneracy of the core-hole with a continuum of states with no core-hole but a high energy Auger electron present. A somewhat different line of application is that of certain rare earth compounds which are known to give rise to mixed valence states\cite{hewson1,hewson2}. These states are characterized by an alternating valence in an otherwise periodic lattice which can result in  unusual thermodynamic properties. The reason is that the difference in valence results in a difference in ionic radii and the extra valence electron thus have a strong coupling to the phonon system. The model designed to capture the effect consists of a localized $f$-state (the extra valence electron) coupled to a continuum of delocalized electrons and a phonon field coupled to the $f$-state. Allthough, there is orders of magnitude differences between typical plasmon and phonon energies the physics in the models are very similar and only the model parameters differ.

Finally the local polaron model has been applied to the problem of resonant tunneling\cite{wingreen} in the context of electronic transport, and the very similar problem of Hot Electron Femtochemistry at Surfaces\cite{gadzuk91} (HEFatS). The idea of HEFatS is that an adsorbate system on a metal surface can have unoccupied electronic states which obtain a broadening due to interaction with the metallic states. If a hot electron (an electron above the Fermi level) is generated in the metal it may interact with the unoccupied state and induce a chemical reaction on the surface. As an example we can think of a single molecule on a metal surface with one unoccupied electronic state well above the Fermi level. A hot electron with an energy that matches the unoccupied orbital has the possibility of tunnelling from the metal to the molecule resulting in a transient occupation of the orbital. If the molecule was initially in an equilibrium position the electron will assert a force on the internal molecular degrees of freedom and can excite vibrational modes of the molecule before it tunnels back into the conductor. The molecule may acquire enough energy in this process to undergo a chemical reaction or a desorption event. A clever method to produce hot electrons is based on a Metal-Insulator-Metal (MIM) heterostructure as suggested by Gadzuk\cite{gadzuk96}. With an ideal MIM device it is possible to tune hot electrons to any desired resonance of an adsorbate system and the approach thereby suggests the highly attractive possibility of performing selective chemistry at surfaces. Such devices have been constructed and characterized\cite{thomsen} and comprise a promising candidate for advanced HEFatS experiments.

Common to all these applications is that a linear bosonic coupling term has been assumed. It is by no means obvious that linear coupling captures the possibly complicated interaction of bosons and electrons, although it is probably often a good approximation. To examine the local polaron model beyond linear coupling we calculate the consequences of substituting the linear coupling term with a quadratic coupling term. In principle we should add the quadratic coupling on top of the linear, but this renders the model somewhat tedious to work with and the physics of quadratic coupling become hidden in complicated expressions. In contrast, having only quadratic coupling allow us to obtain inelastic scattering amplitudes very similar to those with linear coupling and the comparison is very instructive. In terms of bosonic potentials, a linear coupling term corresponds to a shift in the potential minimum whereas a quadratic coupling term corresponds to a shift in the frequency of the potential.

The paper is organized as follows: In section \ref{models} we present the local polaron model with a general coupling function and no bosonic dispersion. The electronic part is briefly reviewed and the wide band limit which is imposed in the remainder of the paper is defined. We then present the well known spectral function and inelastic scattering probabilities of the model with linear coupling and compare with a quadratically coupled model calculated in the present work. It is shown that for inelastic scattering involving a large number of bosons, quadratic coupling can give rise to much larger scattering probabilities. In section \ref{application} we apply the
theory to hot electron mediated excitation of the different modes of CO adsorbed on Cu(100). The model parameters are calculated using density functional theory and the delta self-consistent field method and we find that linear coupling dominates desorption probabilities for the normal modes along the molecular axis, but vanishes for the frustrated rotations where quadratic coupling has to be taken into account. In appendix \ref{decoupling} we derive a path integral representation of the Newns-Anderson retarded Green function and show that the special properties of the wide band limit allow us to decouple bosonic and electronic degrees of freedom. Appendix \ref{linear} and appendix \ref{quadratic} present the details of the calculations leading to the spectral functions and inelastic scattering probabilities associated with linear and quadratic coupling. In appendix \ref{combined}, we show how a linear transformation of creation operators makes it possible to obtain the exact Green functions including both linear and quadratic coupling.

\section{Model}\label{models}
\subsection{The Newns-Anderson model with coupling to bosons} 
The general model we are concerned with is composed of a Newns-Anderson type Hamiltonian\cite{newnsanderson1,newnsanderson2} coupled to a dispersionless (single frequency) boson field through a single electronic state. A dispersionless boson field naturally corresponds to a single mode of oscillation in an adsorbate system, whereas we can think of the dispersionless model as describing an Einstein band if the boson field represents a phonon system. Thus it is a model of non-interacting metallic electrons $|k\rangle$, a localized resonant state $|a\rangle$, and a harmonic oscillator described by the coordinate $x$ or equivalently the bosonic creation and annihilation operators $a^\dag$ and $a$. The Hamiltonian is given by
\begin{align}\label{H}
H=&\sum_k\epsilon_kc_k^{\dag}c_k+\sum_k\Big(V_{ak}c_a^{\dag}c_k+V_{ak}^*c_k^{
\dag}c_a\Big)\notag\\
&+\hbar\omega_0 a^\dag a+\varepsilon_a(x)c_a^{\dag}c_a,
\end{align}
where $c_a$ creates an electron in the state $|a\rangle$ and $c_k$ creates an electron in the state $|k\rangle$. The function $\varepsilon_a(x)$ couples the resonant electron to the oscillator degrees of freedom. If one considers a hole coupled to the bosons instead of an electron, the order of $c_a$ and $c_a^\dag$ should be exchanged.

It is natural to Taylor expand the coupling function in the vicinity of the ground state minimum $x_0$. Including only the zeroth order term $\varepsilon_a(x_0)=\varepsilon_0$ results in the Newns-Anderson model. Since it is quadratic in the electronic creation and annihilation operators one could in principle formally diagonalize it. However, it is much more useful to investigate the resonant state $|a\rangle$ (which is not an eigenstate of the Hamiltonian) and we are thus led to consider the retarded Green function:
\begin{align}
 G_R^0(t)=-i\theta(t)\langle0|c_a(t)c^\dag_a(0)|0\rangle,
\end{align}
where $|0\rangle$ is an electronic vacuum state and 
\begin{equation}
 c(t)=e^{iHt/\hbar}c(0)e^{-iHt/\hbar}.\notag
\end{equation}
It is easily calculated in the energy domain using the Dyson equation and the result is
\begin{align}\label{G_R} 
G_R^0(\omega)=\frac{1}{\hbar\omega-\varepsilon_0-\Sigma(\omega)+i\Gamma(\omega)/2},
\end{align}
with
\begin{align}
\Gamma(\omega)=2\pi\sum_k|V_{ak}|^2\delta(\hbar\omega-\epsilon_k),
\end{align}
and
\begin{align}
\Sigma(\omega)=\int\frac{d\omega'}{2\pi}\frac{\Gamma(\omega)}{\omega-\omega'}.
\end{align}
Assuming the hopping matrix elements $V_{ak}$ to be constant, $\Gamma(\omega)$ becomes proportional to the metal density of states. If we furthermore assume the metal density of states to be wider than the resonance energy we can write $\Gamma(\omega)=\Gamma(\varepsilon_0)$ and $\Sigma=0$. This is the wide band limit which will be imposed in the present paper. It allow us to separate electronic and bosonic degrees of freedom in the general case and we can calculate Green functions corresponding to linear and quadratic coupling exactly.

In the wide band limit the electronic retarded Green function is
\begin{equation}\label{G_0}
 G_R^0(t)=-i\theta(t)e^{-(i\varepsilon_0+\Gamma/2)t/\hbar},
\end{equation}
and the spectral function is a Lorentzian with Full Width at Half Maximum given by $\Gamma$. When boson coupling terms are included (first and second order Taylor expansions of $\varepsilon_a(x)$ in \eqref{H}) the spectral function changes and inelastic scattering on the resonance becomes possible.

Suppose the resonance is initially unoccupied and the oscillator is in the state $n$. The differential probability that an incoming particle (hole or electron) with energy $\varepsilon$ will scatter through the resonance into a state of energy $\varepsilon'$ is given by the inelastic  scattering matrix which can be expressed in terms of a two-particle Green function\cite{wingreen} as
\begin{align}\label{R_inel}
 R(n;\varepsilon',\varepsilon)=&\Gamma^2\int\frac{d\tau dsdt}{2\pi\hbar^3}e^{i(\varepsilon-\varepsilon')\tau/\hbar+i\varepsilon't/\hbar-i\varepsilon s/\hbar}\notag\\
 &\times G(n;\tau,s,t),
\end{align}
where
\begin{align}
G(n;\tau,s,t)=\theta(s)\theta(t)\langle n|c_a(\tau-s)c^\dag_a(\tau)c_a(t)c^\dag_a(0)|n\rangle.\notag
\end{align}
The probability of transferring a given amount of energy to the bosons can thus be obtained by integrating the inelastic scattering matrix over the relevant values of $\varepsilon'$. In this paper we focus on inelastic scattering by electrons, since this is the relevant quantity in the context of HEFatS and EELS. However, the two-particle Green function also appear in the calculation of optical transition amplitudes of an adsorbed molecule\cite{mahan} and knowing $G(n;\tau,s,t)$ allows one to calculate a variety of observable quantities.

Finally, we note that the lifetime of the electron is independent of the boson coupling in the wide band limit. The probability that the state $|a\rangle$ is unoccupied and that the oscillator is in any state at time $t$ given that the state was occupied and the oscillator was in the state $|n\rangle$ at $t=0$ is
\begin{equation}
 p_a(n;t)=\sum_{m=0}^\infty|\langle m,a;t|n,a;0\rangle|^2=e^{-\Gamma t/\hbar},
\end{equation}
which is proved in appendix \ref{decoupling}.
 
\subsection{Coupling function and adiabatic potentials}\label{adiabatic}

\begin{figure}[htp]
	\includegraphics[scale=0.40]{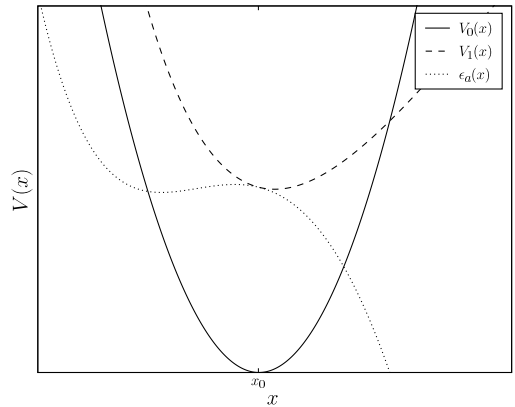} 
\caption{A general example of adiabatic potentials $V_1(x)$ and $V_0(x)$ and the coupling function $\varepsilon_a(a)=V_1(x)-V_0(x)$. The vertical distance between the two potentials at the ground state minimum is $\varepsilon_0$.}
\label{fig:potential}
\end{figure}    	
Consider the state $|x,a\rangle$ with the oscillator at $x$ and an electron occupying the resonance. The expectation value of the Hamiltonian on such a state will depend on the value of $x$ due to the coupling $\varepsilon_a(x)$ and if we could calculate the electronic energy for all values of $x$ we would obtain an excited state potential $V_1(x)=\langle x,a|H|x,a\rangle$. Doing the same for the state with no electron in the resonance $|x,0\rangle$ would result in a different potential $V_0(x)$ and the coupling function should then be given by $\varepsilon_a(x)=V_1(x)-V_0(x)$ which is illustrated in figure \ref{fig:potential}. In the model \eqref{H} we have implicitly assumed that the potential $V_0(x)$ is quadratic, but in general it could have any form. The potentials $V_1(x)$ and $V_0(x)$ are called Born-Oppenheimer surfaces and are obtained by moving the oscillator adiabatically in the electronic environment.

\subsection{Linear coupling}
\begin{figure}[htp]
	\includegraphics[scale=0.40]{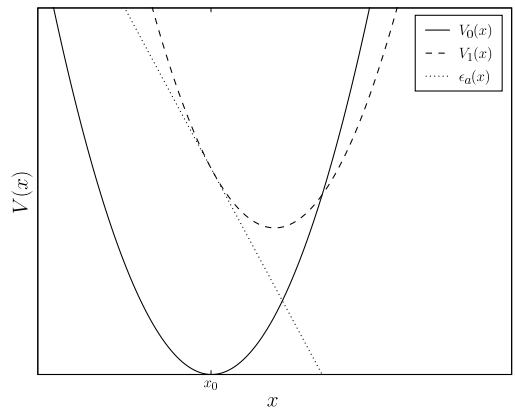} 
\caption{A shifted excited state corresponds to a linear coupling function $\varepsilon_a(x)\sim x$. 
The strength of the coupling is proportional to the derivative of the excited state at $x_0$.}
\label{fig:linear_pot}
\end{figure}
We now Taylor expand the coupling function $\varepsilon_a(x)$ to first order and express the boson coordinate in terms of creation and annihilation operators. This gives an interaction term:
\begin{equation}
 H_I=\lambda_1c_a^\dag c_a(a^\dag+a),
\end{equation}
with
\begin{align}
\lambda_1=\frac{l}{\sqrt{2}}\frac{\partial}{\partial x}V_1\Big|_{x=x_0}, \qquad l=\sqrt{\frac{\hbar}{m\omega_0}},
\end{align}
where $m$ and $\omega_0$ are the mass and frequency of the oscillator. This model corresponds to the potentials and coupling function shown in figure \ref{fig:linear_pot}.

As shown in appendix \ref{decoupling}, the bosonic degrees of freedom decouple from electronic degrees of freedom in the wide band limit and the retarded Green function thus becomes a product of an electronic part given by \eqref{G_0} and an bosonic part. Since the interacting term is linear in the oscillator coordinate the oscillator part of the Hamiltonian can be diagonalized by ``completing the square`` Or, equivalently, performing a canonical transformation which shifts the boson coordinate an amount proportional to $c_a^\dag c_a$:
\begin{equation}
 H\rightarrow e^{iP}He^{-iP},\qquad P=-i\frac{\lambda_1}{\hbar\omega_0}c_a^\dag c_a(a^\dag-a).
\end{equation}
The retarded Green function can then be calculated exactly for the $n$'th excited state giving
\begin{align} 
G^{(1)}_R(n;t)=&-i\theta(t)e^{(-i\varepsilon_0-\Gamma/2)t/\hbar}e^{-g_1(1-i\omega_0t-e^{-i\omega_0 t})}\notag\\
&\times L_n\big[g_1|1-e^{i\omega_0t}|^2\big],
\end{align}
where $L_n$ is the $n$'th Laguerre polynomial and $g_1=\lambda_1^2/(\hbar\omega_0)^2$. In this paper $g_n$ denotes a dimensionless effective coupling constant and $G^{(n)}$ denotes the \textit{exact} Green function corresponding to a coupling term $\varepsilon_a(x)\sim x^n$ and \textit{not} the contribution from an $n$'th order perturbative calculation as is sometimes custom. The spectral function is given by $A_n^{(1)}(\omega)=-2\mathtt{Im}G_R^{(1)}(n;\omega)$ and for the ground state we obtain
\begin{align}\label{linear_spectral}
 A_0^{(1)}(\omega)&=\Gamma e^{-g_1}\\ &\times\sum_{m=0}^\infty\frac{g_1^m}{m!}\frac{1}{(\hbar\omega-\varepsilon_0+(g_1-m)\hbar\omega_0)^2+(\Gamma/2)^2}\notag.
\end{align}
The spectral function is thus a sum of Lorentzians of width $\Gamma$ and an internal spacing of $\omega_0$ and the amplitude of the $m$'th peak follows a Poisson distribution. It should be noted that the peaks do not represent excited states of the oscillator. It is the spectral function of the resonant electron with the oscillator in the ground state and the different peaks show that the coupling term mixes the eigenstates of the isolated oscillator. The real part of the self energy is always negative and given by $-\hbar\omega_0g_1$ and all physical observables are invariant to $\lambda_1\rightarrow-\lambda_1$ since linear coupling corresponds to a shifted harmonic oscillator and the direction of the shift is irrelevant.

The two-particle Green function and inelastic scattering matrix can also be calculated exactly\cite{wingreen, olsen} and the probability that an incoming electron scatters on the resonance and excites the oscillator from the ground state to the $n$'th excited state is
\begin{align}\label{P_1}
P_n^{(1)}(\varepsilon)=\Gamma^2e^{-2g_1}\frac{g_1^n}{n!}|F^{(1)}_n(\varepsilon)|^2,
\end{align}
 with
\begin{align}
F^{(1)}_n(\varepsilon)=&\sum_{k=0}^n(-1)^j\binom{n}{j}\notag\\
\times&\sum_{l=0}^\infty\frac{g_1^l}{l!}\frac{1}{\varepsilon-\varepsilon_0+(g_1-j-l)\hbar\omega_0+i\Gamma/2}\notag.
\end{align}
The probability of exciting the $n$'th vibrational state thus essentially conserves the Poisson distribution, but the Lorentzians are replaced by the interference factor $|F_n(\varepsilon)|^2$. The results \eqref{linear_spectral} and \eqref{P_1} can also be obtained using a disentangling theorem\cite{mitter} as shown in appendix \ref{linear}.
 
\subsection{Quadratic coupling}
\begin{figure}[htp]
	\includegraphics[scale=0.40]{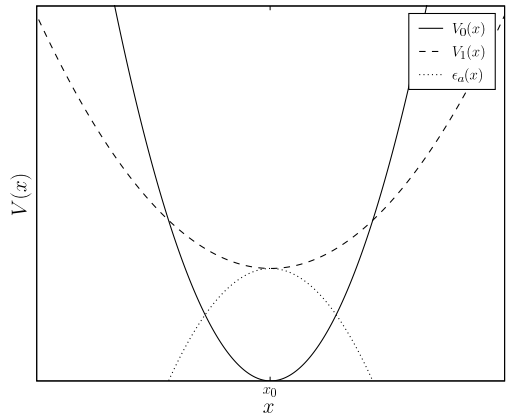} 
\caption{A frequency shifted excited state gives rise to a quadratic coupling function: $\varepsilon_a(x)\sim x^2$.}
\label{fig:quadratic_pot}
\end{figure}
We will now consider an quadratic excited state potential energy surface $V_1(x)$ which has a minimum that coincides with the ground state minimum, but has a different harmonic evolution. The potentials and coupling function corresponding to this is shown in figure \ref{fig:quadratic_pot}. Alternatively we could regard this model as a second order Taylor expansion of the phonon coupling function $\varepsilon_a(x)$ when the first order contribution vanishes. The interaction term in the Hamiltonian becomes
\begin{equation}
 H_I=\lambda_2c_a^\dag c_a(a^\dag+a)^2,
\end{equation}
with
\begin{equation}
\lambda_2=\frac{\hbar}{2m\omega_0}\frac{1}{2}\frac{\partial^2(V_1-V_0)}{\partial x^2}\Big|_{x=x_0}.
\end{equation}
In this work we will only consider bound excited state potentials of the form $V_1(x)=m\omega_1^2(x-x^0)^2/2$ and we can then write
\begin{equation}
\lambda_2=\frac{(\hbar\omega_1)^2-(\hbar\omega_0)^2}{4\hbar\omega_0},\qquad\omega_1=\omega_0\sqrt{1+4\lambda_2/\hbar\omega_0}.
\end{equation}
In the wide band limit the electronic and bosonic degrees of freedom decouple and the boson propagator can be evaluated using a generalization of the Baker-Campbell-Hausdorff formula\cite{mitter}. As a result the retarded Green function and spectral function can be calculated exactly. The derivation is shown in appendix \ref{quadratic} and gives for the ground state:
\begin{align}\label{A_2}
 &A_0^{(2)}(\omega)=\Gamma\sqrt{1-g_2}\\ &\times\sum_{m=0}^\infty\frac{b_mg_2^m}{(\hbar\omega-\varepsilon_0+\hbar(\omega_0-\omega_1)/2-2m\hbar\omega_1)^2+(\Gamma/2)^2}\notag,
\end{align}
where
\begin{equation}
g_2=\Big(\frac{\omega_0-\omega_1}{\omega_0+\omega_1}\Big)^2,\quad b_m=\frac{1}{m!}\frac{\partial^m}{\partial x^m}(1-x)^{-1/2}\Big|_{x=0}.\notag
\end{equation}
This result is valid for $\omega_1>0$ which implies that $g_2<1$. Again, the spectral function is a sum of Lorentzians, but with the $m$'th peak damped by a factor of $b_mg_2^m$ instead of a Poisson distribution. The real part of the self energy is now given by half the frequency shift. The internal spacing between the peaks is $2\omega_1$ and we see that the quadratic coupling only mixes the oscillator ground state with the even excited states of $V_1(x)$. This is due to the mirror symmetry of $\varepsilon_a(x)$ which implies that only oscillator states with equal parity mix. In the quadratic case, the effective dimensionless coupling $g_2$ is not simply given by $\lambda_2^2/(\hbar\omega_0)^2$ as may have been anticipated, and thus the $m$'th term in \eqref{A_2} does not correspond to a $m$'th order perturbative calculation of $G_R^{(2)}(t)$ in $\lambda_2$. The calculation leading to the exact result \eqref{A_2} is very different from the perturbative approach and we have checked that the second order Taylor expansion of \eqref{A_2} indeed gives the result obtained from second order perturbation theory.

Allthough the spectral function \eqref{A_2} shows a series of peaks spaced by $2\omega_1$ it is only possible to excite an integer number of $\omega_0$ through inelastic scattering. The reason is of course that the boson field is completely decoupled from the electronic states in the asymptotics of a scattering event and will thus be observed in an free oscillator eigenstate. Again, the symmetry of the quadratic coupling means that transitions involving an uneven number of bosons are forbidden. The two-particle Green function and inelastic scattering matrix are calculated in appendix \ref{quadratic} and the probability for an incoming hot electron to excite $2n$ quanta of oscillation when initially in the ground state is:
\begin{align}\label{P_2}
P^{(2)}_{2n}(\varepsilon)=\Gamma^2(1-g_2)b_ng_2^n|F^{(2)}_n(\varepsilon)|^2
\end{align}
with
\begin{align}
F^{(2)}_n&(\varepsilon)=\sum_{j=0}^n(-1)^j\binom{n}{j}\sum_{k=0}^\infty\sum_{l=0}^\infty\frac{b_lg_2^{k+l}(n+k-1)!}{k!(n-1)!}\notag\\
&\times\frac{1}{\varepsilon-\varepsilon_0+(\hbar\omega_0-\hbar\omega_1)/2-2(j+k+l)\hbar\omega_1+i\Gamma/2}.\notag
\end{align}
The structure is very similar to the case of linear coupling. With linear coupling the probability for an electron to create $n$ bosons are proportional to the $n$'th order Taylor expansion of $e^{g_1}$ and normalized by $e^{-2g_1}$ whereas in the quadratic case the probability to create $2n$ bosons are proportional to the $n$'th order Taylor expansion of $(1-g_2)^{-1/2}$ and normalized by $(1-g_2)$. In the context of EELS and plasmon excitations, one would now observe a series of peaks spaced by $2\omega_0$. If the plasma frequency is not known the spacing itself cannot give clues to whether linear or quadratic coupling governs the transitions, but one could use the relative amplitude between peaks since these follow a Poisson distribution if linear coupling dominates and the distribution $b_ng_2^n$ if quadratic coupling dominates. If both linear and quadratic coupling is present one would observe a coupling dependent enhancement of every second peak.

\begin{figure}[t]
	\includegraphics[scale=0.3]{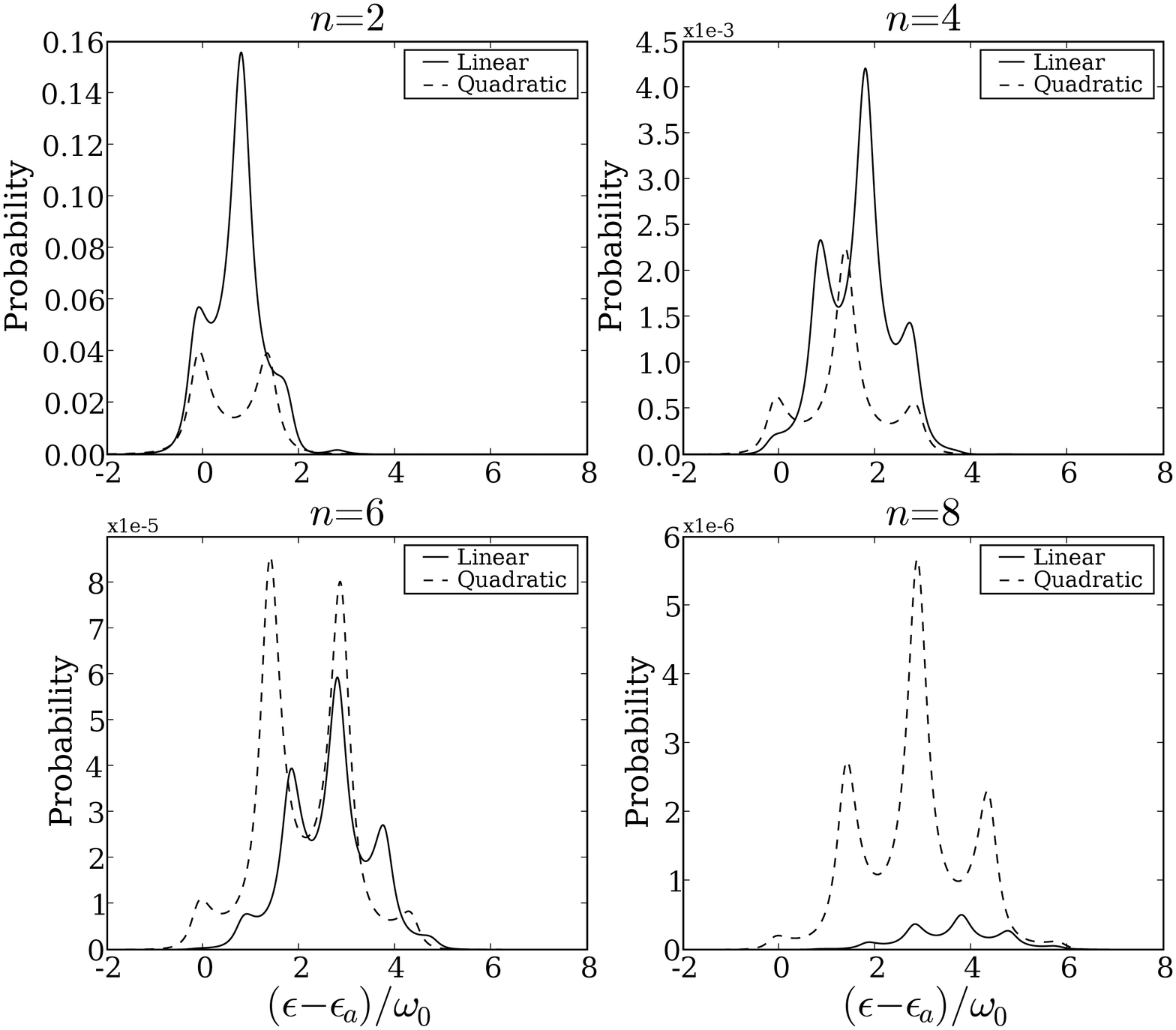} 
\caption{Probabilities of making the transition $0\rightarrow n$ through resonant inelastic scattering with linear and quadratic coupling. The parameters are $\Gamma/\hbar\omega_0=0.5$, $g_1=0.2$, and $\omega_1=0.75\omega_0$ $(g_2=0.02)$. Even though $g_2<g_1^2$ the quadratic coupling becomes dominating for large $n$ due to the slowly decaying expansion coefficients. One should also note the spacing between peaks which is $\omega_0$ for linear coupling and $2\omega_1$ for quadratic coupling. The centers of the probability distributions are approximately shifted by $n\omega_0/2$ for the linear coupling and $n\omega_1/2$ for the quadratic coupling relative to the bare resonance energy $\varepsilon_0$, since this is where the binomial coefficients in \eqref{P_1} and \eqref{P_2} have their maxima.}
\label{fig:prob}
\end{figure}

In a model with linear coupling, the probability of exciting $2n$ vibrational quanta is proportional to $g_1^{2n}$ whereas it is proportional to $g_2^{n}$ in a quadratic coupled model. This implies that if $g_2>g_1^2$ a quadratic coupling term will give rise to larger inelastic scattering probabilities than a linear term. Even with $g_2<g_1^2$ a quadratic coupling term may have stronger effect for large $n$ since the expansion coefficients of $(1-x)^{-1/2}$ decay slower than those of $e^x$. This is illustrated in figure \ref{fig:prob}, where the probability of transferring $n$ vibrational quanta to the ground state is shown for linear and quadratic coupling.

\section{Application to hot electron mediated desorption}\label{application}
As an example of a system where the dynamics can be approximated by a local polaron model, we consider the problem of hot electron mediated energy transfer on a metal surface. Such an energy transfer can lead to desorption of adsorbed molecules \cite{prybyla1,prybyla2,budde,misewich2,struck,howe,cai,fournier} or induce chemical reactions which cannot proceed by thermal heating.\cite{bonn2}. The conceptual picture of the process is the following: Hot electrons are generated in the metal by means of an MIM device or a femtosecond laser. The hot electrons may then interact with a chemisorped molecule by tunneling from the metal to an unoccupied molecular state and excite vibrational states in the molecule. If enough energy is transferred to the molecule either by a single or multiple scattering events, the molecule may eventually desorp or break an internal chemical bond. As a particular example we will calculate transition probabilities for CO adsorbed on Cu(100).

To calculate inelastic scattering probabilities within the local polaron model we need to obtain the coupling function $\varepsilon_a(x)$. As described in section \ref{adiabatic}, we can fix the molecule at different positions and calculate the potential energy surfaces $V_1(x)$ and $V_0(x)$ at each point and $\varepsilon_a(x)=V_1(x)-V_0(x)$. The model \eqref{H} does not directly contain Coulomb interactions between electrons, but these are included in the calculation of $\varepsilon_a(x)$ which thus becomes an effective coupling that is supposed to contain all the electronic interaction associated with the excited state of the molecule. 

The potential energies $V_1(x)$ and $V_0(x)$ has been obtained using the code
\texttt{gpaw}\cite{gpaw, mortensen} which is a real-space Density
Functional Theory (DFT) code that uses the projector augmented wave
method.\cite{blochl1,blochl2} In all our calculations we used the
Revised Perdew-Burke-Ernzerhof (RPBE) exchange-correlation functional
\cite{hammer} since this has been designed to perform well for
molecules adsorbed on surfaces, and has been shown to perform better
than the original PBE functional\cite{PBE} for adsorbed molecules.

We set up a Cu(100) surface consisting of three atomic
layers with the top layer being relaxed. 10 \AA\ of vacuum has then been
introduced above the slab and 0.50 monolayer of adsorbate molecules
relaxed at top sites which is the preferred adsorption site.
Both molecules adsorb with their molecular axis perpendicular to the surface with O pointing away from the surface. 
We then did a normal mode analysis and mapped out the three ground state potential energy functions $V_0(x_i)$ corresponding to the two normal modes that involve the perpendicular degrees of freedom and a frustrated rotation. The perpendicular modes roughly correspond to an internal stretch $d=x_O-x_{C(N)}$ and center of mass $z=(m_Ox_O+m_{C(N)}x_{C(N)})/(m_O+m_{C(N)})$. We do not include the three remaining molecular modes since one is another frustrated rotation with identical properties to the one considered, and the two frustrated translations are only weakly coupled to the resonant electron and are not expected to play a significant role in the femtochemistry. In all calculations we use a p(2x2) cell, sample 12 irreducible k-points in the surface plane, and use a grid spacing of 0.2 {\AA}.

\begin{figure}[t]
	\includegraphics[scale=0.40]{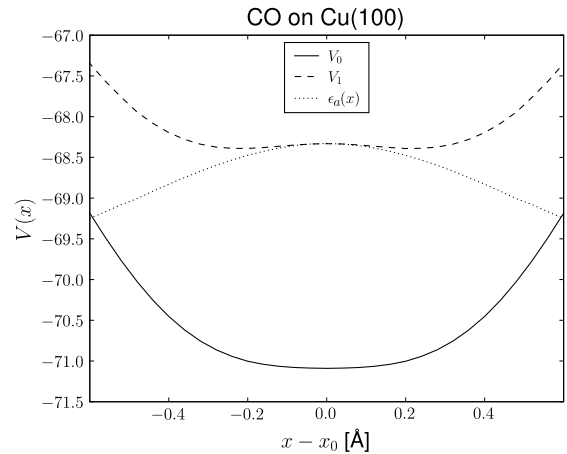} 
\caption{Potential energy surfaces along the frustrated rotation mode of CO adsorbed on a Cu(100) surface. The coordinate $x$ is a generalized coordinate representing the deviation from equilibrium. $x=0.4$ corresponds to a $24^\circ$ angular deviation from the perpendicular position.}
\label{fig:pes}
\end{figure}

To find the excited state potential energies $V_1(x_i)$ corresponding to the three normal modes of interest, we have used the method
of linear expansion $\Delta$SCF which has been published in a previous
work\cite{gavnholt} and implemented in \texttt{gpaw}.  In the previous
publication we have tested the method against inverse
photo-emission spectroscopy, and found that it performed well for
molecules chemisorped on surfaces.\cite{gavnholt} In each step of the
self consistency cycle an electron is removed from the Fermi level,
the density of an excited state is added to the total density, and the
band energy of this state is added to the total energy. To get the
band energy right we need to expand the excited state on the Kohn-Sham
orbitals found in each iteration. The method is thus a generalization
of the usual $\Delta$SCF where occupation numbers are
changed. Instead of changing occupation numbers we occupy an orbital
which is not an eigenstate of the Kohn-Sham Hamiltonian but a superposition
of eigenstates in such a way that the state is as close as possible to
the original molecular state. In the present case the excited state is the anti-bonding $2\pi$ orbital of CO. In figure \ref{fig:pes} we show the ground and excited state potential energy surfaces corresponding to the frustrated rotation along with $\varepsilon_a(x)$. It is clear that the excited state potential is not exactly a quadratic potential and the parameter $\lambda_2$ which we need to calculate transition probabilities will depend on how we fit this potential to a quadratic form. However the width of the Gaussian ground state vibrational wavefunction corresponds to $x=0.08$ {\AA} and for low lying excitations we can thus use this region of the potential which is rather flat. In fact, a closer look at the excited state potential reveals that the ground state minimum geometry actually has an unstable extremum in the excited state, but since the curvature is rather small we will simply approximate it by a constant potential. For both perpendicular modes we find that $\varepsilon_a(x_i)\sim x_i$ and quadratic coupling can thus be neglected. In contrast, due to symmetry the excited state potential energy of frustrated rotation is invariant to $x_i\rightarrow-x_i$ and the linear coupling term thus vanishes. We have calculated the excitation energy to $\varepsilon_a(x_0)=2.8\;eV$ and the resonance width is estimated from the Kohn-Sham projected density of states to $\Gamma\approx1.0\;eV$. In table \ref{tab}, we display the calculated parameters corresponding to the three modes. 
\begin{table}[t]
\begin{center}
\begin{tabular}{c|c|c|c}
Mode & $\hbar\omega$ & $\lambda_1$ & $\lambda_2$ \\
	\hline
Frustrated rotation  & 0.037 & 0      & -0.009 \\
Center of mass       & 0.043 & -0.006 & $\sim0$ \\
Internal stretch     & 0.248 & -0.170 & $\sim0$
\end{tabular}
\end{center}
\caption{Parameters for CO adsorbed on Cu(100). All number are $eV$. Note that while the quadratic coupling for the two perpendicular modes are very small and thus neglectable, the linear coupling of frustrated rotation vanishes exactly due to symmetry.}
\label{tab}
\end{table}

\begin{figure}[b]
	\includegraphics[scale=0.45]{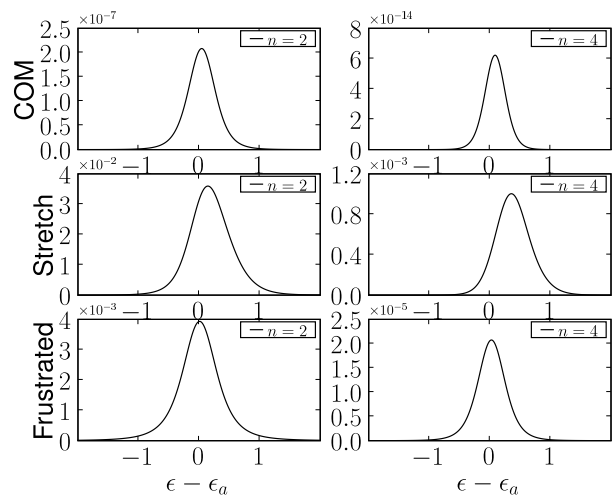} 
\caption{Probabilities of exciting two and four quanta of vibrations to the center of mass, internal stretch and frustrated rotation modes CO adsorbed on Cu(100).}
\label{fig:CO_prob}
\end{figure}

We note that when calculating transition probabilities we should include all modes in the model \eqref{H}, because even if the modes are not coupled directly they have an indirect coupling since they all interact with the resonance. It is possible to obtain expressions for the scattering matrix including more than one mode, but these are rather complicated to handle and for weakly coupled systems the physics can usually be extracted from three one-mode models.\cite{olsen} In figure \ref{fig:CO_prob} we show the calculated probabilities for a hot electron to excite the different modes of CO adsorbed on Cu(100). The internal stretch and and frustrated rotation show transition probabilities on the same order of magnitude whereas the center of mass vibrations are very unlikely to get excited. This is in accord with calculations of the electronic friction coefficients of this system\cite{head-gordon,tully} which is very closely related to the coupling function $\varepsilon_a(x)$.\cite{brandbyge}
The frequency of internal vibration is five times larger than both the center of mass and frustrated rotation frequencies and as previously shown\cite{olsen} the stretch mode will completely dominate the total energy transfer. Thus, in a simple model where hot electron mediated desorption\cite{struck} is reduced to calculating the probability of transferring the chemisorption energy to the adsorbate, the internal mode governs the desorption probability. Nevertheless, our estimate of $\Gamma$ is based on the Kohn-Sham density of states which may give a poor description of the electronic spectral function $A^0(\omega)$. If $\Gamma$ is significantly smaller than our estimate, the quadratically coupled frustrated rotation will play an important role in hot electron mediated desorption for this system.

\section{Summary and discussion}\label{summary}
We have calculated the spectral function and inelastic scattering amplitudes in a local polaron model with quadratic coupling to bosons. The probability of exciting $n$ bosons is found to be damped by a distribution function given by the $n$'th Taylor expansion of $1/\sqrt{1-g_2}$ which decays much slower than the Poisson distribution appearing in a linearly coupled model. Hence for comparable values of linear and quadratic coupling constants, a quadratic term will dominate inelastic scattering probabilities involving a large number of bosonic excitations.

As an application we have considered the problem of hot electron mediated vibrational excitations of molecules adsorbed on metal surfaces. The coupling constants were calculated from the excitation energy along the molecular normal modes using delta self-consistent field DFT. It was found that quadratic coupling is important for exciting the frustrated rotations since this mode does not couple linearly due to symmetry.

A major approximation in the model is the quadratic assumption for the ground state potential. In our numerical example with HEFatS
it is clear from figure \ref{fig:pes} that the potentials is not exactly quadratic. For the center of mass mode the situation is even worse and a Morse potential is much better suited to describe this mode. The anharmonic deviations are likely to have a significant effect on high lying excited states, but renders the model much more complicated. In fact, since the coupling to the internal stretch mode seems to govern the rate of energy transfer,\cite{olsen} one has to assume that the energy is readily redistributed to other degrees of freedom and anharmonic coupling is thus expected to play a vital role in the actual desorption process. 

The wide band limit has been essential in the derivation of scattering amplitudes, and we do not have the means to solve the model (with linear or quadratic coupling) exactly beyond this approximation. However, it would be very interesting to do perturbation theory with the general retarded Green function \eqref{G_R} to examine the effect of energy dependence in the electronic self energy.

\begin{acknowledgments}
We are grateful to Karsten Jacobsen for his advice and help with the path integral representation of the Newns-Anderson model and to Jakob Schi\o tz for helpful comments and guidance into the exciting field of Hot Electron Femtochemistry. The Center for Individual Nanoparticle Functionality (CINF) is sponsored by the Danish National Research Foundation. This work was supported by the Danish Center for Scientific Computing.
\end{acknowledgments}

\begin{appendix}
\begin{widetext}
\section{Decoupling of electronic and bosonic degrees of freedom in the wide band limit}\label{decoupling}
\subsection{Path integral representation of the Newns-Anderson retarded Green function}
The path integral representation of propagators often give a renewed insight into the underlying physics, allthough the mathematical complexity can be somewhat larger. By writing the Newns-Anderson retarded Green function as a sum over paths, we see that each path can be understood as sequence of jumps from the resonance to the metallic band and we then have to sum over all possible time intervals between each jump. In the wide band limit the time spend in the metal band goes to zero and the electron thus spends all the time of propagation in the resonant state. The Newns-Anderson model is given by \eqref{H} with $\varepsilon_a(x)=\varepsilon_a(x_0)=\varepsilon_0$:
\begin{align}\label{H_0}
 H_0=\varepsilon_0c_a^\dag c_a+\sum_k\epsilon_kc_k^\dag c_k+\sum_k\Big(V_{ak}c_a^\dag c_k+V_{ak}^*c_k^\dag c_a\Big)
\end{align}
and
\begin{align}
 G_R^0(t)=-i\theta(t)\langle a|e^{-iHt}|a\rangle.
\end{align}
The path integral representation is derived by dividing the time interval $t$ in $N$ intervals of length $\Delta t=t/N$. When $N$ becomes sufficiently large we can take $e^{-iHt}=(e^{-iH\Delta t})^N\approx(1-iH\Delta t)^N$. We then insert $N-1$ complete sets of states $|n\rangle$ such that the Green function becomes a sum over $N$-fold products of matrix elements
\begin{align}\label{G_sum}
 G_R^0(t)\approx-i\theta(t)\sum_{n_1,n_2\ldots n_{N-1}}\langle a|1-iH_0\Delta t|n_1\rangle\langle n_1|1-iH\Delta t|n_2\rangle\ldots\langle n_{N-1}|1-iH_0\Delta t|a\rangle.
\end{align}
Assuming that $\langle a|k\rangle=0$ the states $|n\rangle$ can either be $|a\rangle$ or $|k\rangle$ and the matrix elements $\langle a|1-iH_0\Delta t|a\rangle=e^{-i\varepsilon_0\Delta t}$, $\langle k_1|1-iH_0\Delta t|k_2\rangle=\delta_{k_1k_2}e^{-i\epsilon_k\Delta t}$ and $\langle a|1-iH_0\Delta t|k\rangle=-iV_{ak}\Delta t$ represent propagation in the resonance, propagation in the band and a jump from band to the resonance respectively. When we take the limit $N\rightarrow\infty$, \eqref{G_sum} becomes formally exact and the jumps between band and resonance become instantaneous. It is then most convenient to order the terms in \eqref{G_sum} according to the number of jumps. Since the endpoints of the time interval is at the resonance, a jump into the band has to be accompanied by a jump back into the resonance and each such "band excursion" comes with a factor of $-\sum_k|V_{ak}|^2e^{-i\epsilon_k\tau_i}$ where $\tau_i$ is the time spend in the $i$'th excursion into band. It is also clear that $p$ excursions into the resonance has to be accompanied by $p+1$ resonant propagation factors $e^{-i\varepsilon_0\sigma_i}$ where $\sigma_i$ is the $i$'th time interval in the resonant state. Finally, for a given number of band excursion we have to integrate over all possible band and resonance time intervals and the retarded Green function becomes
\begin{align}\label{G_path1}
 G_R^0(t)=&-i\int_0^\infty d\sigma_0e^{-i\varepsilon_0\sigma_0}\sum_{p=0}^{\infty}\bigg(-\int_0^\infty\int_0^\infty d\sigma d\tau\sum_k|V_{ak}|^2e^{-i\epsilon_k\tau}e^{-i\varepsilon_0\sigma}\bigg)^p\notag\\
 &\times\delta\Big(\sigma_0+\sum_j(\sigma_j+\tau_j)-t\Big),
\end{align}
where the delta function has been introduced to ensure that the time intervals sum to $t$ and the theta function has become redundant. We can use the delta function to eliminate the $\sigma_i$ integration variables and get
\begin{align}\label{G_path2}
G_R^0(t)=\int\mathcal{D}\chi e^{iS_0(\chi)}=&-ie^{-i\varepsilon_0t}\int_0^\infty d\sigma_0\sum_{p=0}^{\infty}\bigg(-\int_0^\infty\int_0^\infty d\sigma d\tau\Gamma(\tau)\bigg)^p\delta\Big(\sigma_0+\sum_j(\sigma_j+\tau_j)-t\Big),
\end{align}
with
\begin{align}\label{gamma}
\Gamma(t)\equiv\sum_k|V_{ak}|^2e^{-i(\epsilon_k-\epsilon_0)t}=\int_{-\infty}^\infty\frac{d\omega}{2\pi}\Gamma(\omega)e^{-i(\omega-\varepsilon_0)t}.
\end{align}
By using that $\delta(t)=(1/2\pi)\int d\omega e^{i\omega t}$ it is now possible to evaluate \eqref{G_path2} and recover the result \eqref{G_R}. In the wide band limit, $\Gamma(t)=\Gamma\delta(t)$, which implies that the electron does not spend any time in the band and the retarded Green function becomes a sum over paths which are composed of \textit{instantaneous} excursions into the band. We use the notation $\chi$ to represent a position in state space and $\int\mathcal{D}\chi e^{iS_0(\chi)}$ as a formal expression representing the sum over all paths weighted by the Newns-Anderson action $S_0$.

\subsection{Resonant electron in a bosonic environment}
We now proceed with the full Hamiltonian \eqref{H}. Introducing the bosonic coordinate $x$ the full retarded Green function (with the boson field in the state $x_0$) can be written
\begin{align}
G_R(x_0;t)=\int\mathcal{D}\chi\mathcal{D}x e^{iS_0(\chi)+iS_B(x)+iS_I(\chi,x)},
\end{align}
where $S_0(\chi)$ is the Newns-Anderson action, $S_B(x)$ is the free bosonic action corresponding to the Hamiltonian $H_B=\omega_0a^\dag a$, and $S_I$ is the interaction part of the action corresponding to the Hamiltonian $H_I=c_a^\dag c_a\varepsilon_a(x)$. However, there is a much nicer way to handle the coupling to the boson field. One can think of the bosons as an environment influencing the paths of the resonant electron and it can be shown that the Green function can be written\cite{hedegard1,hedegard2}
\begin{align}
G_R(x_0;t)=\int\mathcal{D}\chi e^{iS_0(\chi)}\langle x_0|\widetilde{U}(\chi;t)|x_0\rangle,
\end{align}
with the environment time evolution operator
\begin{align}
\widetilde{U}(\chi;t)=e^{iH_Bt}\mathcal{T}e^{-i\int_0^tdt'\widetilde{H}_I(\chi(t'))},
\end{align}
where $\widetilde{H}_I(\chi(t'))=H_B+H_I(\chi(t'))$ is the environment Hamiltonian evaluated on an electronic state fixed at $\chi$ and $\mathcal{T}$ denotes time ordering. Thus, the price we pay in separating bosonic and electronic degrees of freedom is an explicit path dependence in the environment part of the propagator. In general it is not possible to evaluate $\widetilde{H}_I(\chi(t'))$ on all possible paths, but in the wide band limit it is particularly simple. The reason is that the resonant electron stays on the resonance in all possible paths and the environment Hamiltonian is therefore \textit{independent} of the electronic path. In fact, the environment propagator becomes $\widetilde{U}(\chi;t)=e^{i\omega a^\dag at}e^{-i\omega a^\dag at-i\varepsilon_a(x)t}$ and the electronic and bosonic degrees of freedom completely decouple in the retarded Green function:
\begin{align}
G_R(n;t)=G_R^0(t)G_B(n;t),\qquad G_B(n;t)=\langle n|e^{i\omega a^\dag at}e^{-i\omega a^\dag at-i\varepsilon_a(x)t}|n\rangle.
\end{align}
The situation is very similar for the two-particle Green function:
\begin{align}
G(n;\tau,s,t)=\theta(s)\theta(t)\langle n|c_a(\tau-s)c^\dag_a(\tau)c_a(t)c^\dag_a(0)|n\rangle=\theta(s)\theta(t)\langle n|\hat c_a(\tau-s)U(\tau-s,\tau)\hat c^\dag_a(\tau)\hat c_a(t)U(t,0)\hat c^\dag_a(0)|n\rangle.\notag
\end{align}
The resonant electron is first propagated forward in time from $0$ to $t$ and then backward in time from $\tau$ to $\tau-s$. The interaction vanishes between the $c^\dag(\tau)$ and $c(t)$ because the resonant state is unoccupied here. Again, the full Green function can be written in terms of a bosonic influence propagator and in the wide band limit the electronic and bosonic degrees of freedom decouple so:
\begin{align}
G(n;\tau,s,t)=G^0_R(t)\bar G^0_R(s)G_B(n;\tau,s,t)
\end{align}
with
\begin{align}
G_B(n;\tau,s,t)=\langle n|e^{iH_0(\tau-s)}e^{i(H_0+\varepsilon_a(x))s}e^{-iH_0\tau}e^{iH_0t}e^{-i(H_0+\varepsilon_a(x))t}|n\rangle.
\end{align}
This can be seen by applying the arguments above to both of the time evolution operators and the fact that $G^0(\tau,s,t)=G^0_R(t)\bar G^0_R(s)$. 

We observe that for any coupling function $\varepsilon_a(x)$ the above form of the two-particle Green function implies that $G(n;t,t,t)=|G^0_R(t)|^2$. This means that in the wide band limit, the resonant lifetime is unaffected by the phonon coupling since the probability of finding the electron in the state $|a\rangle$ at time $t$ (and the oscillator in any state $|m\rangle$) given that it was there at $t=0$ (where the oscillator was in the state $|n\rangle$) is
\begin{align}
p_a(n;t)&=\sum_{m=0}^\infty|\langle m,a;t|n,a;0\rangle|^2=\sum_{m=0}^\infty|\langle m|c_a(t)c_a^\dag|n\rangle|^2=\langle n|c_ac_a^\dag(t) c_a(t)c_a^\dag|n\rangle\notag\\
&=G(n;t,t,t)=|G^0_R(t)|^2=e^{-\Gamma t}.
\end{align}
So, in the wide band limit the resonant state always has a well defined lifetime given by $T_a=\hbar/\Gamma$

The problem of calculating the inelastic scattering matrix has now been reduced to evaluating the phonon propagator $G_B(n;\tau,s,t)$. In general this is not an easy task, but we will show that in the case of linear and quadratic coupling terms we can use a disentangling theorem\cite{mitter} to write the exponential operators in a form that allows a direct evaluation of the expectation value. The theorem is a generalization of the Baker-Campbell-Hausdorff theorem and states that if $A$, $B$, and $C$ are three operators with a closed commutator algebra then $e^{aA+bB+cC}=he^{\alpha A}e^{\beta B}e^{\gamma C}$, where $h$, $\alpha$, $\beta$, and $\gamma$ are known functions of $a,b,c$, and the commutation parameters.

\section{Green functions with linear coupling to bosons}\label{linear}
\subsection{Single-particle Green function}
The model \eqref{H} with linear coupling function given by 
\begin{align}
 \varepsilon_a(x)=\lambda_1(a^\dag+a),
\end{align}
is well-known and the one-particle Green functions can be obtained exactly in the wide band limit by a canonical transformation.\cite{mahan, wingreen} Here we derive it using the formalism above and the disentangling theorem.\cite{mitter} To obtain the one-particle Green function we need to evaluate the boson propagator
\begin{align}
 G^{(1)}_B(n;t)=\langle n|e^{i\omega_0ta^\dag a}e^{-i\omega_0t(a^\dag a+\lambda_1(a+a^\dag)/\omega_0)}|n\rangle.
\end{align}
Using the disentangling theorem on the second exponential operator leads directly to the expression
\begin{align}
 G^{(1)}_B(n;t)=&e^{in\omega_0t}e^{-g_1(1-i\omega_0t-e^{-i\omega_0t})}\langle n|e^{-\frac{\lambda_1}{\omega_0}(1-e^{-i\omega_0t})a^\dag}e^{\frac{\lambda_1}{\omega_0}(1-e^{i\omega_0t})a}e^{-i\omega_0ta^\dag a}|n\rangle\notag\\
 =&e^{-g_1(1-i\omega_0t-e^{-i\omega_0t})}L_n\big[g_1|1-e^{i\omega_0t}|^2\big],\qquad g_1=\Big(\frac{\lambda_1}{\omega_0}\Big)^2
\end{align}
where $L_n(x)$ is the $n$'th Laguerre polynomial. To obtain the ground state spectral function \eqref{linear_spectral} we Taylor expand $\exp(g_1e^{i\omega_0t})$, perform a Fourier transformation, and take the imaginary part.

\subsection{Two-particle Green function}
The procedure can also be used to obtain the two-particle Green function. The object of interest is now the two-particle boson propagator which we write
\begin{align}
 G^{(1)}_B(n;\tau,s,t)=&\langle n|e^{i\omega_0(\tau-s)a^\dag a}e^{i\omega_0s(a^\dag a+\lambda_1(a+a^\dag)/\omega_0)}e^{-i\omega_0\tau a^\dag a}e^{i\omega_0ta^\dag a}e^{-i\omega_0t(a^\dag a+\lambda_1(a+a^\dag)/\omega_0)}|n\rangle\notag\\
 =&e^{in\omega_0(\tau-t)}e^{-g_1(1-i\omega_0t-e^{-i\omega_0t})}e^{-g_1(1+i\omega_0s-e^{i\omega_0s})}\notag\\
 &\times\langle n|e^{\frac{\lambda_1}{\omega_0}(1-e^{-i\omega_0s})a^\dag}e^{-\frac{\lambda_1}{\omega_0}(1-e^{i\omega_0s})a}e^{i\omega_0(t-\tau) a^\dag a}e^{-\frac{\lambda_1}{\omega_0}(1-e^{-i\omega_0t})a^\dag}e^{\frac{\lambda_1}{\omega_0}(1-e^{i\omega_0t})a}|n\rangle,
\end{align}
where we used two different forms of the disentangling theorem to move $e^{i\omega_0sa^\dag a}$ and $e^{-i\omega_0ta^\dag a}$ to the left and right respectively. We then use the theorem again to move $e^{i\omega_0(t-\tau)a^\dag a}$ to the left and obtain
\begin{align}
G^{(1)}_B(n;\tau,s,t)=&\notag e^{-g_1(2-i\omega_0(t-s)-e^{-i\omega_0t}-e^{i\omega_0s})}\\
&\times\langle n|e^{\frac{\lambda_1}{\omega_0}(1-e^{-i\omega_0s})e^{-i\omega_0(t-\tau)}a^\dag}e^{-\frac{\lambda_1}{\omega_0}(1-e^{i\omega_0s})e^{i\omega_0(t-\tau)}a}e^{-\frac{\lambda_1}{\omega_0}(1-e^{-i\omega_0t})a^\dag}e^{\frac{\lambda_1}{\omega_0}(1-e^{i\omega_0t})a}|n\rangle.\notag
\end{align}
Finally, we can use the Baker-Campbell-Hausdorff theorem to collect all lowering operators at the the right. The evaluation of the remainder gives a Laguerre polynomial and the result is
\begin{align}
G^{(1)}_B(n;\tau,s,t)=e^{ig_1\omega_0(t-s)}e^{-g_1f_{\omega_0}(\tau,s,t)}L_n\big([g_1(f_{\omega_0}+f^*_{\omega_0})\big],
\end{align}
with
\begin{align}
f_{\omega_0}(\tau,s,t)=2-e^{-i\omega_0t}-e^{i\omega_0s}+e^{-i\omega_0\tau}(1-e^{i\omega_0t})(1-e^{i\omega_0s}).
\end{align}
The inelastic scattering matrix has previously been derived\cite{wingreen,olsen} and we will not repeat the calculation here.

\section{Green functions with quadratic coupling to bosons}\label{quadratic}
\subsection{Single-particle Green function}
We now consider a quadratic coupling function of the form
\begin{align}
 \varepsilon_a(x)=\lambda_2(a^\dag+a)^2
\end{align}
To obtain the retarded Green function we would like to calculate the boson propagator
\begin{align}
G^{(2)}_B(n;t)=\langle n|e^{i\omega_0ta^{\dag}a}e^{-i\omega_0 a^{\dag}at-i\lambda_2(a^{\dag}a+aa^{\dag}+aa+a^\dag a^\dag)t}|n\rangle=e^{ni\omega_0t}e^{-i\lambda_2t}\langle n|e^{-i(\omega_0+2\lambda_2)a^{\dag}at-i\lambda_2(aa+a^\dag a^\dag)t}|n\rangle.
\end{align}
We proceed by disentangling the exponential operator:
\begin{align}\label{two_particle_dis}
e^{-i(\omega_0+2\lambda_2)a^{\dag}at-\lambda_2(aa+a^\dag a^\dag)t}=e^{i\lambda_2t}e^{i\omega_0/2t}e^{g/2}e^{fa^\dag a^\dag}e^{fe^{-2g}aa}e^{ga^\dag a},
\end{align}
where
\begin{align}
f(t)=\frac{-\lambda_2\tanh(i\omega_1t)}{\omega_1+(\omega_0+2\lambda_2)\tanh(i\omega_1t)},\qquad g(t)=-\ln\Big(\cosh(i\omega_1t)+\frac{\omega_0+2\lambda_2}{\omega_1}\sinh(i\omega_1t\Big),
\end{align}
and $\omega_1=\omega_0(1+4\lambda_2/\omega_0)^{1/2}$. This is valid for a bound excited state potential with positive second derivative in which case the argument of the square root is positive. In the case of an unbound excited state potential, the functions $f$ and $g$ involve real hyperbolic functions and the spectral function acquires a qualitatively different structure. Acting with the operator $e^{\alpha aa}$ on a state $|n\rangle$ gives
\begin{align}
 e^{\alpha aa}|n\rangle=\sum_{l=0}^{[n/2]}\frac{\alpha^l}{l!}\Big(\frac{n!}{(n-2l)!}\Big)^\frac{1}{2}|n-2l\rangle,
\end{align}
where $[n/2]$ means the integer part of $n/2$. Collecting it all and noting that $f(t)e^{-g(t)}=\frac{-\lambda_2}{\omega_1}\sinh(i\omega_1t)$, gives the retarded Green function in the wide band limit:
\begin{align}\notag
 G^{(2)}_R(n;t)=-i\theta(t)e^{(-i\varepsilon_0-\Gamma/2)t}e^{i(n+1/2)\omega_0t}\Big(\cosh(i\omega_1t)+\frac{\omega_0+2\lambda_2}{\omega_1}\sinh(i\omega_1t\Big)^{-n-\frac{1}{2}}\sum_{l=0}^{[n/2]}\frac{h^{2l}}{(l!)^2}\frac{n!}{(n-2l)!}
\end{align}
with
\begin{align}
 h=\frac{\lambda_2}{\omega_1}\sinh(i\omega_1t)
\end{align}
To find the spectral function of the oscillator ground state we Taylor expand the square root and obtain
\begin{align}\label{taylor}
G^{(2)}_R(n=0;t)&=-i\theta(t)e^{(-i\varepsilon_0-\Gamma/2)t}e^{i\omega_0t/2}e^{-i\omega_1t/2}\Big(\frac{\omega_1+\omega_0+2\lambda_2}{2\omega_1}\Big)^{-\frac{1}{2}}\Big(1-\frac{\omega_0-\omega_1+2\lambda_2}{\omega_0+\omega_1+2\lambda_2}e^{-2i\omega_1t}\Big)^{-\frac{1}{2}}\notag\\
&=-i\theta(t)e^{(-i\varepsilon_0-\Gamma/2)t}e^{i\omega_0t/2}\Big(\frac{\omega_1+\omega_0+2\lambda_2}{2\omega_1}\Big)^{-\frac{1}{2}}\sum_{m=0}^{\infty}b_mg_2^me^{-i(2m+1/2)\omega_1t},\qquad\omega_1>0
\end{align}
with
\begin{align}
g_2=\frac{\omega_0-\omega_1+2\lambda_2}{\omega_0+\omega_1+2\lambda_2}=\Big(\frac{\omega_0-\omega_1}{\omega_0+\omega_1}\Big)^2,\qquad b_m=\frac{1}{m!}\frac{\partial^m}{\partial x^m}(1-x)^{-1/2}\Big|_{x=0}.
\end{align}
Fourier transforming and taking the imaginary part then gives
\begin{align}
A_0^{(2)}(\omega)=\Gamma\sqrt{1-g_2}\sum_{m=0}^\infty\frac{b_mg_2^m}{(\omega-\varepsilon_0+(\omega_0-\omega_1)/2-2m\omega_1)^2+(\Gamma/2)^2},\qquad\omega_1>0
\end{align}
where we also used that $2\omega_1/(\omega_1+\omega_0+2\lambda_2)=1-g_2$. Note that the condition of $\omega_1>0$ implies that $g_2<1$. 

\subsection{Two-particle Green function}
We now need the propagator
\begin{align}\label{G_2}
G^{(2)}_B(n;\tau,s,t)=&\langle n|e^{i\omega_0a^{\dag}a(\tau-s)}e^{i(\omega_0+2\lambda_2)a^{\dag}as+i\lambda_2(aa+a^\dag a^\dag)s} e^{-i\omega_0a^{\dag}a\tau}e^{i\omega_0a^{\dag}at}e^{-i(\omega_0+2\lambda_2)a^{\dag}at-i\lambda_2(aa+a^\dag a^\dag)t}|n\rangle\notag\\
=&e^{in\omega_0(\tau-s)}\sum_{m=0}^{\infty}e^{im\omega_0(t-\tau)}\langle n|e^{i(\omega_0+2\lambda_2)a^{\dag}as+i\lambda_2(aa+a^\dag a^\dag)s}|m\rangle\langle m|e^{-i(\omega_0+2\lambda_2)a^{\dag}at-i\lambda_2(aa+a^\dag a^\dag)t}|n\rangle.
\end{align}
We restrict the calculation to the ground state two-particle Green function which using the disentangled expression \eqref{two_particle_dis} becomes
\begin{align}
G^{(2)}(n=0;\tau,s,t)=G^0_R(t)\bar G^{0}_R(s)e^{i\omega_0(t-s)/2}e^{g(t)/2+g(-s)/2}\sum_{m=0}^\infty e^{2im\omega_0(t-\tau)}\frac{f^m(t)f^m(-s)(2m)!}{(m!)^2}.
\end{align}
It is also possible to obtain a closed expression that does not involve the infinite sum, since instead of inserting a complete set in \eqref{G_2} we could have brought all lowering operators to the left by repeated use of the disentangling theorem. However, to calculate the inelastic scattering matrix \eqref{R_inel} we need to integrate over $\tau$ which is more convenient in the present form. Performing the $\tau$ integral and using the resulting delta function to replace $2m\omega_0$ with $\varepsilon-\varepsilon'$, leaves the two remaining integrals as complex conjugates. We note that $b_m=(2m)!/4^m(m!)^2$ and write
\begin{align}
R^{(2)}(\varepsilon', \varepsilon)=\Gamma^2\sum_{m=1}^\infty4^mb_m\delta(\varepsilon-\varepsilon'-2m\omega_0)|D_m(\varepsilon)|^2,
\end{align}
with
\begin{align}
D_m=\int_0^\infty dte^{-i(\varepsilon_0-\varepsilon-\omega_0/2-i\Gamma/2)t}\Big(\cosh(i\omega_1t)+\frac{\omega_0+2\lambda_2}{\omega_1}\sinh(i\omega_1t)\Big)^{-1/2}\bigg(\frac{-\lambda_2\tanh(i\omega_1t)}{\omega_1+(\omega_0+2\lambda_2)\tanh(i\omega_1t)}\bigg)^m.\notag
\end{align}
The reason we have excluded the $m=0$ term is that it does not give rise to inelastic scattering and the elastic part of the scattering matrix have an additional term that we do not consider here.\cite{wingreen} This expression implies that quadratic coupling can only give rise to inelastic scattering events involving an even number of vibrational quanta. This is also true if the initial state is not the ground state, since from \eqref{G_2} we see that in general $(m-n)$ has to be even. To evaluate the inelastic scattering matrix we note that
\begin{align}
\bigg(\frac{-\lambda_2\tanh(i\omega_1t)}{\omega_1+(\omega_0+2\lambda_2)\tanh(i\omega_1t)}\bigg)^m&=\Big(\frac{-\lambda_2}{\omega_1+\omega_0+2\lambda_2}\Big)^m\Big(1-e^{-2i\omega_1t}\Big)^m\Big(\frac{1}{1-g_2e^{-2i\omega_1t}}\Big)^m\notag\\
&=\frac{g_2^{m/2}}{2^m}\sum_{j=0}^m(-1)^j\binom{m}{j}e^{-2ij\omega_1t}\sum_{k=0}^\infty\frac{(m+k-1)!}{k!(m-1)!}g_2^ke^{-2ik\omega_1t},\qquad\omega_1>0\notag
\end{align}
where we used that $-\lambda_2/(\omega_1+\omega_0+2\lambda_2)=\sqrt{g_2}/2$. The Taylor expansion of the square root gives
\begin{align}
\Big(\cosh(i\omega_1t)+\frac{\omega_0+2\lambda_2}{\omega_1}\sinh(i\omega_1t\Big)^{-1/2}=\sqrt{1-g_2}\sum_{l=0}^{\infty}b_lg_2^le^{-i(2l+1/2)\omega_1t},\qquad\omega_1>0.
\end{align}
leading to
\begin{align}
D_m=&\frac{i\sqrt{1-g_2}g_2^{m/2}}{2^m}\sum_{j=0}^m(-1)^{j}\binom{m}{j}\sum_{k=0}^{\infty}\sum_{l=0}^{\infty}\frac{b_lg_2^{k+l}(m+k-1)!}{k!(m-1)!}\notag\\
&\times\frac{1}{\varepsilon-\varepsilon_0+(\omega_0-\omega_1)/2-2(j+k+l)\omega_1+i\Gamma/2},
\end{align}
and
\begin{align}
R^{(2)}(\varepsilon', \varepsilon)&=\Gamma^2(1-g_2)\sum_{m=1}^\infty b_mg_2^m\delta(\varepsilon-\varepsilon'-2m\omega_0)\notag\\
&\times\bigg|\sum_{j=0}^m(-1)^j\binom{m}{j}\sum_{k=0}^{\infty}\sum_{l=0}^{\infty}\frac{b_lg_2^{k+l}(m+k-1)!}{k!(m-1)!}
\times\frac{1}{\varepsilon-\varepsilon_0+(\omega_0-\omega_1)/2-2(j+k+l)\omega_1+i\Gamma/2}\bigg|^2\notag.
\end{align}

\section{Linear and quadratic coupling combined}\label{combined}
It is in principle straightforward to generalize the expressions above to the case of a linear \textit{and} a quadratic coupling term in the Hamiltonian. The linear term can be transformed away by noting that
\begin{align}
\omega_0a^\dag a+\lambda_1(a^\dag+a)+\lambda_2(a^\dag+a)^2=\omega_0\tilde a^\dag\tilde a+\lambda_2(\tilde a^\dag+\tilde a)^2-\gamma\lambda_1,
\end{align}
with
\begin{align}
\tilde a=a+\gamma,\qquad\tilde a^\dag=a^\dag+\gamma,\qquad\gamma=\frac{\lambda_1}{\omega_0+4\lambda_2},\qquad[\tilde a,\tilde a^\dag]=1.
\end{align}
Since the commutator algebra of $\tilde a$ and $\tilde a^\dag$ is identical to that of $a$ and $a^\dag$ we can immediately write down the one-particle boson propagator in its disentangled form
\begin{align}
G_B^{(1,2)}(n;t)=e^{i\gamma\lambda_1t}e^{i(n+1/2)\omega_0t}e^{g/2}\langle n|e^{f\tilde a^\dag\tilde a^\dag}e^{g\tilde a^\dag \tilde a}e^{f\tilde a\tilde a}|n\rangle
\end{align}
Using that 
\begin{align}
e^{g\tilde a^\dag \tilde a}=e^{-\gamma^2(1-e^g)}e^{\gamma(e^g-1)a^\dag}e^{-\gamma(e^{-g}-1)a}e^{ga^\dag a},
\end{align}
we can evaluate the propagator in the ground state and obtain
\begin{align}
G_B^{(1,2)}(n=0;t)=e^{i\gamma\lambda_1t}e^{i\omega_0t/2}e^{g/2}e^{-\gamma^2(1-e^g-2f)}.
\end{align}
The expression clearly reduces to $G_B^{(1)}(t)$ and $G_B^{(2)}(t)$ in the limits $\lambda_1\rightarrow0$ and $\lambda_2\rightarrow0$ respectively. It should be straightforward to obtain the spectral function by Fourier transforming this expression after a Taylor expansion of the exponentials. However, the result becomes rather involved and we will not attempt to do the calculation here.
\end{widetext}
\end{appendix}

\newpage


\end{document}